\documentclass[12pt]{article} 
\usepackage[hyperfootnotes=false]{hyperref}
\usepackage{amsmath}
\usepackage{amssymb}
\usepackage{graphicx}
\usepackage{xcolor}
\newlength{\abstractwidth}
\usepackage{mciteplus} 
\usepackage{subcaption}
\usepackage{bbold}
\usepackage{slashed}

\newcommand{\be}{\begin{equation}}
\newcommand{\bea}{\begin{eqnarray}}
\newcommand{\eea}{\end{eqnarray}}
\newcommand{\beq}{\begin{equation}}
\newcommand{\ee}{\end{equation}}
\newcommand{\eeq}{\end{equation}}

\def\la{\label}

\def\32{{3 \over 2 } }

\def\ba{\begin{eqnarray}}
\def\ea{\end{eqnarray}}

\def\simleq{\; \raise0.3ex\hbox{$<$\kern-0.75em
      \raise-1.1ex\hbox{$\sim$}}\; }
\def\simgeq{\; \raise0.3ex\hbox{$>$\kern-0.75em
      \raise-1.1ex\hbox{$\sim$}}\; }

\def\nnref#1{(\ref{#1})}

\begin{document}

\begin{titlepage}
  \bigskip

  \bigskip\bigskip

  \bigskip

\begin{center}
 
\centerline
{\Large \bf {Black hole entropy and quantum mechanics}}
 \bigskip

 \bigskip
{\Large \bf { }} 
    \bigskip
\bigskip
\end{center}

  \begin{center}

 \bf {Juan Maldacena   }
  \bigskip \rm
  
\bigskip
 Institute for Advanced Study,  Princeton, NJ 08540, U.S.A.

\bigskip
\bigskip

  \end{center}

 \bigskip\bigskip
  \begin{abstract}

We give a brief  overview of black hole entropy,  covering a few main developments since Bekenstein's original proposal\footnote{Contribution to the Jacob Bekenstein memorial volume.}.

 \medskip
  \noindent
  \end{abstract}
\bigskip \bigskip \bigskip

\vspace{1cm}

\vspace{2cm}

  \end{titlepage}



\section{Black hole entropy} 

The idea that black holes have an entropy proportional to the area has been a remarkable realization \cite{Bekenstein:1973ur}. 
It is a surprising confluence of statistical mechanics, quantum mechanics, special  relativity and gravity. 
This can be seen clearly by restoring all the constants normally set to one in the black hole entropy formula \cite{Bekenstein:1973ur,Hawking:1974sw}
\be \la{BHen}
S = { ({\rm Area} )_{\rm horizon} \over 4 l_{\rm Planck}^2 }   ~~~ \longrightarrow~~~ S = k { ({\rm Area } ) c^3  \over 4 G_N \hbar } 
\ee
The formula is most remarkable because a black hole is just a solution of the  Einstein equations, which are classical\footnote{See \cite{Jacobson:2018nnf} for an 
interesting historical and conceptual discussion.}.
This formula, together with Einstein equations is consistent with the first law of thermodynamics  \cite{Bardeen:1973gs}
\be \la{FirstL}
  \delta M = T \delta S   
\ee
where $M$ is the mass of the black holes and $T$ is the temperature of the black hole, given by Hawking's formula \cite{Hawking:1974sw}. 
The fact that the area of the horizon increases \cite{Hawking:1971tu} is consistent with the $2^{nd}$ law of thermodynamics \cite{Bekenstein:1974ax}.
 It also shows
that \nnref{BHen} is a coarse grained entropy, which is the kind
of entropy that increases under the second law. 
 
 These results show that  black holes behave as ordinary thermal  systems for an observer who
 remains outside the black hole.  They further suggest that in the full quantum theory, an outside observer, could view the black hole as an ordinary quantum system. Finding ways to realize this idea was a central topic of research in the 
 past forty years. 
  
Bekenstein proposed a generalized second law for the total entropy, given by the area of the horizon plus the entropy outside the horizon \cite{Bekenstein:1974ax}
\be \label{Sgen}
S_{\rm total} = { ({\rm Area } ) \over 4 l^2_{\rm Planck} } + S_{\rm matter ~outside}
\ee 
The validity of   this generalized second law seems to imply some constraints on the entropic content of matter \cite{Bekenstein:1980jp}. 
This arises from a thought experiment involving  
 matter falling  into a black hole. If we call $S_m $ the entropy of the in-falling matter,  we find that the
 entropy of the outside decreased by $S_m$. However, the area of the black hole increased, as implied by \nnref{FirstL},     $\delta S_{BH} =   \delta E_m/T$.
 So the generalized second law of thermodynamics would be 
 obeyed if this later quantity is larger than $S_m$. This implies that there should be an upper bound on the entropy content in terms of the energy. 
   Matter with some entropy should necessarily carry some energy. 
 Now, this should hold for matter that is close to the black hole horizon. An important property of black holes is that 
 the asymptotic time translation symmetry 
 acts as a boost in the near horizon region. Namely, near the horizon we can define coordinates $x^\pm$ where the radial and time direction has a usual 
 flat space geometry $ds^2 = - dx^+ dx^- + \cdots $. The horizons are at $x^\pm =0$ (past and future horizons). The asymptotic time translation symmetry 
 that shifts the asymptotic time $t$ acts as a boost in the $x^\pm$ plane. We can normalize the boost generator so that $x^+$ has eigenvalue one. 
 In terms of Rindler coordinates, $x^\pm = \pm r e^{ \pm \tau }$ the boost generator is conjugate to $\tau$ time translations. The time $\tau$ is related to the 
 asymptotic time $t$ by 
 \be \la{HakF}
 t = { 1 \over 2 \pi T } \tau 
 \ee
 where $T$ is the Hawking temperature. The precise rescaling is obtained from the full black hole geometry and \nnref{HakF} is then the expression for the Hawking
 temperature in terms of the black hole parameters.
  In other words, in terms of the time $\tau$ the inverse  Hawking temperature is $2 \pi$, so that the physical value of the temperature 
  measured at infinity comes purely
 from the rescaling of the two time coordinates. 
 The same rescaling factor appears in the energy measured with respect to time $t$ versus the boost energy measured with respect to time $\tau$
 \be
 { E \over T} =    2 \pi B  ~,~~~~~~~~~E = i \partial_t ~,~~~~~~B = i \partial_\tau 
 \ee 
 where $B$ is the energy measured with respect to time $\tau$,  and is equal to the boost generator near the horizon. 
 Then the necessary condition for obeying the second law is that 
 \be \la{Boost}
  S_m \leq 2 \pi  B_m 
  \ee
  where $B_m$ is the boost eigenvalue of the matter.
 Bekenstein's formulation of the bound involved bounding the entropy in terms of the energy and the size $R$ of the 
 system \cite{Bekenstein:1980jp}. However, \nnref{Boost} is a more precise form of the bound, it is what can be proven and it is also what we need for the second law. 
 
 It was suspected for many years that \nnref{Boost} would imply some restriction on the types of QFT that can be coupled to gravity. However, it was shown by
  Casini \cite{Casini:2008cr}, building on previous ideas by \cite{Marolf:2003wu,Marolf:2003sq}, that a version of \nnref{Boost} is actually true in
   {\it any} relativistic quantum field theory.  
 
 Before discussing the proof, it is necessary to notice that  \nnref{Boost} has an $\hbar$ in the left hand side. So the bound is
 trivially obeyed in the classical limit. In order to get close to saturating the bound it is necessary to consider quantum effects. 
 In relativistic quantum field theory  we cannot localize  particles or excitations. Therefore, in order to talk about a bound like 
 \nnref{Boost} we need to define things more precisely. In fact, the proper quantum version involves the Von Neumann entropy of the quantum field theory 
 state restricted to the right Rindler wedge. This Von Neumann entropy is sometimes called ``entanglement'' entropy since it arises because we are dividing a pure state into two parts, one outside  the horizon and the other inside. 
   We can compute this entropy for the state in question, containing some matter, and we subtract the same quantity for the vacuum to form the difference $\Delta S$. 
 Then the bound is \cite{Casini:2008cr}
 \be \label{CasB}
 \Delta S \leq 2 \pi \langle \Delta B \rangle 
 \ee
 where $\Delta B$ is the  
  expectation value of the Boost generator in the state in question minus its expectation value in the vacuum. There are divergencies when we compute the entropies
  but these cancel when we compute the differences. There can be some remaining ambiguities in the definition of the entropy but these are the same as the 
  ambiguities defining the precise form of the boost generator on the half space. 
 It turns out that \nnref{CasB} is the same as the positivity of relative entropy \cite{Casini:2008cr}. Equivalently, we can think of the 
 vacuum as  thermal state on the Rindler wedge. Then  \nnref{CasB} reduces to the statement  that 
  the difference in free energy between the state in question and  the vacuum   is positive, which is true since the thermal equilibrium state minimizes the free energy. 
   
   Another set of ideas that grew out of Bekenstein's bound is summarized in \cite{Bousso:2018bli}, and will not be discussed here. 
   
  In recent years the discussion of black hole thermodynamics has been extended to the semiclassical 
  domain, where we consider also quantum fields and include their entropy. 
  In that case,  a more precise version of black hole entropy includes also the Von Neumann entropy of the quantum fields outside the black hole horizon \cite{Bombelli:1986rw} 
  \be \la{FullEnt}
  S_{\rm total}  = { ({\rm Area}) \over 4 G_N } + S_{V.N.} + \cdots
  \ee
  where the dots are some extra pieces that we will not discuss in detail here, having to do with extra Wald terms \cite{Wald:1993nt} and counterterms.
  This a more precise version of \nnref{Sgen}. It is the correct expression up to order $G_N^0$ in the semiclassical expansion. 
    Note that while \nnref{BHen} depends only on gravity, \nnref{FullEnt} depends also on the type of matter fields that we have in the theory. 
  Using the monotonicity of relative entropy, A. Wall, has argued \cite{Wall:2011hj} that the second law of black hole thermodynamics holds for \nnref{FullEnt}. 
 This shows that no new inequality is needed from matter other than the ones that follow from the fact that matter is described by relativistic quantum field theory. 
  
   The fact that black hole entropy exists, and obeys the second law, can be viewed as evidence that black holes should have a unitary description as viewed from the 
outside. Otherwise, if unitarity were preserved only after adding a second asymptotic region beyond the horizon, then why couldn't entropy simply disappear into that
region?

  \section{The search for   black hole microstates}
  
  We can  think of the formula for black hole entropy as a kind of ``experimental'' result, analogous to the entropy that 
  an experimentalist would measure on a real world material by staring from zero temperature and putting in energy slowly to get it to a desired final sate, and computing the entropy from the first law. Since the work of Boltzmann we expect that entropy should be associated to the number of microstates of the system. So the question is: what are the microstates that give rise to the entropy of black holes?. 
   One would expect that these microstates should be a fairly universal since,  at leading order, the entropy  is independent of the details of the matter theory. 
   
   The most naive picture for these microstates comes from simply computing the Von Neumann entropy of the quantum fields outside the black hole. This gives a UV divergent quantity going like \cite{Bombelli:1986rw} 
   \be \la{VN}
   S_{V.N.} = { ({\rm Area} ) \over \epsilon^2 } + \cdots 
   \ee
   where $\epsilon$ is the UV cutoff. We can view this as the entropy of the hot atmosphere near the black hole horizon. 
    In the semiclassical approximation we should always choose the cutoff 
   $\epsilon $ to be larger  than $l_{Planck}$. Therefore \nnref{VN} is smaller than \nnref{BHen}. However, this has not prevented speculation that perhaps by some suitable principle we would find a cutoff procedure that would reproduce
   the black hole entropy. The most precise ideas in this direction involve using a Pauli-Villars type regulator for the fields and viewing the Newton constant as arising as
   in  induced gravity, see e.g. \cite{Demers:1995wg}. This has the drawback that the UV sensitive terms come
   from ghosts. 
   A more correct and solid point of view is to think of \nnref{VN} as part of a correction to black hole entropy and view the 
   divergent term as a term that is cancelled by a similar renormalization of the Newton constant, which leads to a second area term included in the dots in 
   \nnref{FullEnt} that cancels the divergent terms in \nnref{VN}. This cancellation is automatic when we compute the free energy and the entropy using euclidean
    methods \cite{Gibbons:1976ue}. But this still leaves us without a concrete picture for the microstates. Nevertheless it suggests that
   they are related to states that exist near the horizon and near the UV of the bulk. These are states that do not have an explicit, calculable description, within the regime that gravity 
   is a good approximation. Of course, the black hole entropy formula still somehow ``knows'' about them. 

   
      
            \subsection{Black hole entropy in string theory }
   
   \subsection{ Perturbative string theory}
   
   The above discussion suggests that in a UV finite theory of gravity perhaps the Von Neumann entropy is finite and it accounts for the full entropy of the black hole. 
   String theory is definitely UV finite \cite{Witten:2012bh}. Unfortunately, 
    it is not know how to compute the Von Neumann entropy of a subregion. It had been speculated that 
   perhaps the black hole entropy is related to the entropy of open strings ending on the horizon \cite{Susskind:1994sm}. Though this is an attractive idea,
    to the best of my knowledge, there is
   no precise computation of the entropy along these lines. However, there are two more indirect approaches 
    to compute black hole entropy in string theory which we discuss below. 
   
   \subsubsection{Black hole entropy for supersymmetric black holes} 
   
   A very fruitful set up involves the computation of the entropy of extremal   black holes in supersymmetric gravity theories arising from string theory 
   compactifications \cite{Strominger:1996sh}. 
    Extremal black holes are a type of charged black hole. 
  A non-singular charged black hole obeys the condition $M\geq Q$ (in some units). 
  An extremal black hole obeys $M=Q$. In sufficiently  supersymmetric theories
  this sometimes coincides with a BPS bound \cite{Kallosh:1992ii}. A BPS bound is a bound that arises from the supersymmetry algebra, and it also has the form $M\geq Q$. 
  States saturating this bound are specially protected, in the sense that supersymmetry implies that 
   their number  does not change when we change the coupling constants of the theory.  
  By changing the coupling we can start with a black hole solution and turn it into in a weakly coupled collection of strings and D-branes \cite{Polchinski:1995mt}. 
  We will not describe in detail what D-branes are. The important point is that they  are objects that  obey  clear and simple rules. 
  These objects can be assembled in  a large number of quantum states, 
  set by their total charge. For large charges, this number agrees precisely with the one expected from the 
  Bekenstein-Hawking black hole entropy formula \cite{Strominger:1996sh}. 
   
  This counting has been done with increasing degree of precision. Quantum corrections as in \nnref{FullEnt}
have been matched. Most recently it was understood how to perform the gravity computation in a non-perturbative fashion in order to reproduce the precise counting of microstates \cite{Dabholkar:2011ec}. See \cite{Sen:2014aja} for a review and further references. 
Of course, such computations depend on the details of the theory because the entropy of black holes does indeed depend on the details when once we go
 beyond the area term. 

  This has been a great success and it gives us confidence that string theory is a fully consistent theory of quantum gravity.

  \subsubsection{Black hole entropy from the  gauge/gravity duality} 
      
 Black hole entropy can also be studied from the point of view of the gauge/gravity duality \cite{Maldacena:1997re,Gubser:1998bc,Witten:1998qj}. 
 This is a relationship that equates gravity in asymptotically anti-de-Sitter ($AdS$) boundary conditions with a strongly coupled conformal field theory. 
 A black hole in $AdS$ is equated with a thermal state
 in the conformal field theory. This state is a hot fluid of strongly interacting particles. 
The entropy of the black hole is equated to  the entropy of this fluid in the quantum field theory.
 In general, this is difficult to compute   because the field theory is strongly coupled. A particularly simple case is the three dimensional case,  $AdS_3$.
  Here the boundary is $1+1$ dimensional and the powerful conformal symmetry can be used to compute the thermal entropy as \cite{Cardy:1986ie}
 \be
 S = { \pi \over 3}  c L T 
 \ee
 where $L$ is the size of the boundary region, $T$ is the temperature and $c$ is the so called
 central charge of the two dimensional conformal field theory. 
 This central charge can be computed also in the gravity theory \cite{Brown:1986nw}, 
 \be
 c =   { 3 R_{AdS_3} \over 2 G_{N \, 3} } 
 \ee
 So, one can 
  start from the gravity theory, compute the central charge, assume that the theory has a unitary realization as a CFT and then compute the black hole 
  entropy \cite{Strominger:1997eq}. 
 
 Another approach has been to compute the entropy (or energy) numerically in the quantum 
 mechanical description \cite{Berkowitz:2016jlq}. This was done for an  interacting matrix quantum mechanics theory  \cite{Banks:1996vh} that has a gravity dual
 \cite{Itzhaki:1998dd}. This has the advantage of being a direct computation, without any reliance on conformal symmetry. In fact, this quantum mechanical model
 is not conformal invariant.  
 
A feature of these calculations is that they provide the entropy of the full spacetime, including both the black hole and the full spacetime outside. As presently 
understood, these formulas do not allow a clean separation between the black hole degrees of freedom and those of the outside region.

    \section{ Fine grained gravitational entropy } 
    
\begin{figure}[h]
\centerline{\includegraphics[width=14cm]{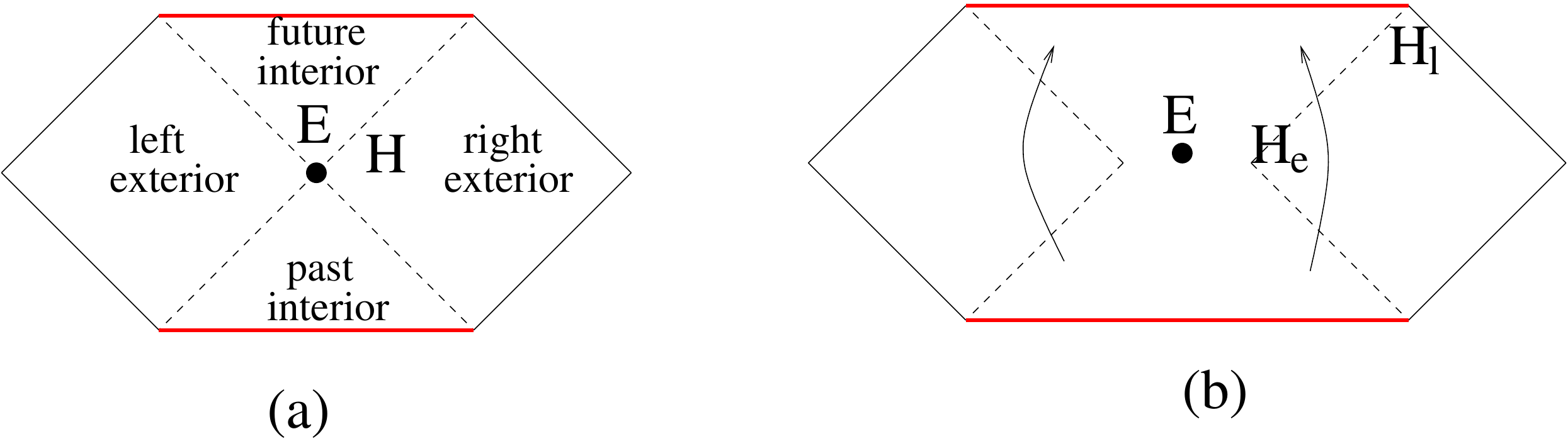}}
\caption{ (a) The Penrose diagram of the eternal Schwarzschild solution. It has the geometry of a non-traversable wormhole connecting two regions that look like the exterior geometry of a black hole. These regions are sharing the interior regions. In this case,  the extremal surface $E$ sits at the event horizon $H$, whose area is
time independent. (b) We consider a configuration where we add extra matter falling into the black holes, represented here by the arrows.
 Now the event  horizon area is a minimum at $H_e$ and grows towards 
its late times $H_l$, with $A_{H_l} > A_{H_e}$ as implied by the area theorem. The extremal surface is inside the wormhole and has a smaller area  
$A_{E} < A_{H_e} < A_{H_l}$.  }
\label{Wormholes}
\end{figure}

   As we mentioned before,  
   the entropy computed by the Bekenstein-Hawking formula should be interpreted as a  coarse grained entropy, since it grows under time evolution,  
  obeying the second law of thermodynamics. 
    One can wonder whether there is any formula that computes the full von Neumann entropy of the 
    microscopic density matrix. In fact, such a  formula was proposed by 
   Ryu and Takayanagi   \cite{Ryu:2006bv} and Hubeny- Rangamani-Takayanagi \cite{Hubeny:2007xt} (HRT). 
   Their prescription is to find an extremal surface in the whole geometry. A surface which has minimal area  on a given spacelike 
    slice but has maximal area among all possible spacelike 
   slices \cite{Wall:2012uf}. This prescription becomes particularly  interesting for spacetimes containing
  a black hole which is  connected through the interior to a second asymptotic region. An example is the 
   standard eternal Schwarzschild black hole, which is 
    the maximal analytic extension of the Schwarzschild solution containing two asymptotic regions, see figure \ref{Wormholes}(a). 
   These configurations can be viewed as entangled black holes \cite{Israel:1976ur,Maldacena:2001kr}. In this case the extremal surface is the bifurcation surface and it
   sits at  the horizon. A more interesting case arises when we consider a deformation of this geometry obtained by adding matter that falls into the two black holes, 
   see figure \ref{Wormholes}(b). In this case the Penrose diagram becomes more elongated and the extremal surface area is smaller than the area of the horizon. Furthermore, the extremal surface is a property of the full spacetime geometry. Therefore, it is time 
   independent,  a fact which is related to the constancy of the fine grained entropy of the system under unitary evolution. 
    An interesting aspect of the extremal surface (or HRT) surface is that it typically  sits behind the event horizon, see figure \ref{Wormholes}(b).

   The HRT   formula is even more remarkable that the Bekenstein-Hakwing formula because it knows about the microscopic entropy of the system. In other words, one might think that gravity is ``just'' a hydrodynamic or thermodynamic approximation to the exact system. However, the HRT formula shows that  gravity knows about the microscopic von Neumann entropy of the system.   
   In the context of holography, the HRT formula can be used to compute the entanglement entropy of subregions of the boundary quantum field theory. 
  This has been useful for two reasons. First, it has provided a tool for studying the entanglement patterns of strongly coupled field theories. The lessons from these
  holographic theories serve as interesting examples in order to understand the problem in more general theories. Second, the fact that the entropy is given by a simple geometric quantity suggests that entanglement plays a crucial role in determining the geometry of spacetime.

\section{Conclusions} 

Black hole entropy has been a central object of interest in the exploration of quantum aspects of gravity. The generalized entropy was defined in the quantum theory 
 and  the generalized second law was proven. This proof works for any matter theory described by relativistic quantum field theory, no further constraint on matter
 was necessary. In particular, a version of the Bekenstein bound \nnref{CasB} is automatically obeyed for any quantum field theory. 

Through string theory we can now identify the black hole microstates. There are two broad ways to do this. One involves supersymmetric black holes, where one can count the states by going to a weakly coupled limit without losing any states. The other involves black holes in special spaces, such as anti-de-Sitter spacetimes, where one can find a dual formulation in terms of a strongly interacting theory on the boundary. In this case,  the black hole looks like a hot fluid on the boundary theory
and the entropy is the entropy of this fluid. This describes the entropy of both the black hole and the spacetime around it. This boundary description is manifestly unitary. So, if we believe the duality we conclude that black holes preserve information. 

The entropy defined in terms of    the area of the event  horizon   is a coarse grained entropy.
 A fined grained notion of entanglement entropy was also recently defined  \cite{Ryu:2006bv,Hubeny:2007xt} and 
has become a very useful window into the quantum mechanics of spacetime.

An important open problem  is to describe the black hole interior in the full quantum theory. In other words, we would like to be able to describe the interior using the same variables in which we describe the microstates with their unitary dynamics. In fact, several paradoxes arise in some naive approaches to this problem 
\cite{Mathur:2009hf,Almheiri:2012rt}.

{\bf Acknowledgements} 

This is a contribution to the {\it Jacob Bekenstein Memorial Volume}. 
It would
  would not have been possible without 
E. Rabinovici's unrelenting insistence.  Other related 
contributions include \cite{Bousso:2018bli,Jacobson:2018nnf}. This work was supported in part by U.S. Department of Energy grant
de-sc0009988 and the ``It from Qubit'' grant from the Simons foundation.
   
\bibliographystyle{utphys}
\bibliography{Bekenstein}{}

\end{document}